\documentclass[aps,prd,amsmath,onecolumn]{revtex4}

\usepackage{epsfig}
\usepackage{graphics}
\usepackage{latexsym}
\usepackage{amsmath}
\usepackage{amssymb}
\usepackage{rotating}
\usepackage{subfigure}
\usepackage{bm}
\usepackage{color}
\usepackage{ulem}

\usepackage[colorlinks = true,
linkcolor = magenta,
urlcolor  = blue,
citecolor = red,
anchorcolor = blue]{hyperref}

\begin{document}
	\title{Probing Lorentz Symmetry Violation via Hadroproduction in Electroweak Pair Annihilation
	}
	
	\author{S.~Mohammad Moosavi Nejad$^{a}$}
	\email{mmoosavi@yazd.ac.ir}
	\author{ Vahid Ekraminasab$^{a}$}
	
	\affiliation{$^{(a)}$Faculty of Physics, Yazd University, P.O. Box
		89195-741, Yazd, Iran}

	\date{\today}
	
	\begin{abstract}
		
		The violation of Lorentz symmetry is anticipated in some theories beyond the Standard Model so  various processes have been studied to search for the effects of Lorentz symmetry violation  (LSV). High energy collision experiments provide a suitable environment where Lorentz symmetry breaking can be examined. In this work, we study the effect of Lorentz-violating terms on the hadron production process from pair annihilation, including a vectorial nonminimal coupling between the fermions and gauge fields in the Electroweak sector.  To be specific, we study the effect of CPT-odd nonminimal Lorentz symmetry violating couplings on the production of bottom flavored mesons through pair annihilation in the Electroweak theory, i.e.,  $e^+e^-\to q\bar{q}\to B+Jets$. 
		Considering the available data and the analytical result for the tree-level  cross section $\sigma^{LO}(e^+e^-\to B+Jets)$ in the presence of LSV terms, we determine an upper bound  on the purely time-like background 4-vector as strict as $\le 10^{-3} GeV^{-1}$, in agreement with previous works on similar Lorentz violating couplings.

	\end{abstract}

	%	\pacs{13.87.Fh, 14.20.Lq, 12.38.Bx, 14.65.Dw.}

	\maketitle

	\section{Introduction}
	\label{sec:intro}
	
	The Standard Model (SM) of particle physics is a successful theory   which describes all known elementary particles and their fundamental interactions,  ranging from electromagnetic to nuclear interactions.
	The symmetries of Gauge and CPT (Charge conjugation-Parity-Time reversal)  as well as the Lorentz invariance  have been the foundation of this basic model from the beginning. We remind that, the Lorentz symmetry is an important feature of nature which refers to the point that experimental results are independent of the orientation or boost velocity of the laboratory through space while  the CPT theorem  predicts the equality of some quantities such as life-time, mass, gyromagnetic ratio and charge-to-mass ratio, for particles and corresponding  antiparticles.
	Despite of many noteworthy achievements, the SM does not show itself as a theory of everything because  there are several observations that cannot be described within its framework, e.g.,  it does not contain the theory of gravity, and does not have convincing answers for the candidates of dark 	matter, matter-antimatter asymmetry, hierarchy problem, etc. This means that the SM cannot be a final theory and it is necessary to go beyond it to achieve a deeper understanding of the laws of nature. To solve the mentioned problems, many new models  and extended theories have been developed during past decades.  
	Among all, a fundamental theory that unifies the SM and gravity would emerge at energies approaching Planck scale ($\approx 10^{19}$~GeV). In the foundation of theories unifying gravity with quantum physics, tiny violations of the Lorentz and CPT symmetries could emerge \cite{Kostelecky:1994rn,Kostelecky:1991ak}. In this context,  the Lorentz violations are allowed in string theory, supersymmetry and Horava-Lifshitz gravity \cite{Horava:2009uw}. To study Lorentz symmetry violations (LSVs) in the context of quantum field theory (QFT)
	a new theory has been constructed by Colladay and Kostelecky \cite{Colladay:1996iz,Colladay:1998fq} which is conventionally called the SM Extension (SME). It  contains the SM, General Relativity and all possible operators breaking Lorentz symmetry.  Generally, effective field theories (EFTs) such as the SME,  introduce Lorentz and CPT violations through spontaneous symmetry breaking caused by hypothetical background fields.
   The structure of SME is one predicted way to study  the Lorentz violation  so that an alternative procedure, which is adopted in our work to study the pair annihilation process in the Electroweak  theory, is to modify just the SME interactions part  via   nonminimal couplings. It leads to the improved  gauge-fermion vertices in the Electroweak (EW) theory.  In practice, through this approach a nonminimal coupling term is added to the covariant derivative which may be CPT-odd or CPT-even. 
    Heretofore, using this prescription  the possibility of Lorentz covariance breakdown in the context of QFT has been extensively studied in the literature. For example, in Ref.~\cite{Charneski:2012py} this possibility was employed to determine the LSV bounds on Bhabha scattering process (a QED process) and in Ref.~\cite{Belich:2006pi} the same procedure has been used for studying the spectrum of hydrogen atom to specify the magnitude  of bounds on  the LSV coefficients. 
    Simultaneously with the theoretical developments, many experimental tests on LSV corrections have also been carried out and several constraints on the LSV parameters have been established, e.g., hyperfine structure of  muonium ground state \cite{Bluhm:1999dx},  clock-comparison experiments \cite{Kostelecky:1999mr},  hyperfine spectroscopy of hydrogen and antihydrogen \cite{Bluhm:1998rk}, etc., see also Refs.~\cite{Mattingly:2005re,ParticleDataGroup:1996dqp,Schwingenheuer:1995uf,NA31:1990xkc}. The clock anisotropy,  which is a spectroscopic experiment, is one of the most precise experiments  \cite{Brown:2010dt} where  the LSV parameters are introduced as in the SME theory. It should be also noted that, the SME includes a number of possible terms which  violate
	local Lorentz invariance by coupling to the spin of particle \cite{Kostelecky:1999mr}.  Then, in Ref.~\cite{Brown:2010dt} authors have  performed a search for neutron spin coupling to a Lorentz- and CPT-violating background field
	using a magnetometer with overlapping ensembles of $K$ and $^3H$ atoms. Therefore, they have determined bounds of $10^{-33}$~GeV  on the LSV parameters. 
	In a realistic theory, if the spontaneous CPT and partial LSV are extended to the four-dimensional spaces, detectable effects might occur in interferometric experiments with neutral kaons \cite{Kostelecky:1991ak,Kostelecky:1994rn}, neutral  $B_s$- or $B_d$-mesons \cite{Colladay:1994cj,Kostelecky:1996fk}, or neutral D-mesons \cite{Colladay:1994cj,Kostelecky:1996fk,Colladay:1995qb}. For example, the quantities parameterizing indirect CPT violation in these systems could be nonzero.\\
	As was mentioned, various processes have been yet considered to study the LSV 	effects but there are few
	studies about possible effects of LSV on scattering processes, see for example the study on Bhabha scattering in Refs.~\cite{Charneski:2012py,Casana:2012vu}.
	Historically, hadron and lepton colliders have played complementary roles in high-energy physics. In this context, high energy collision experiments provide a suitable 	situation where the LSV effects can be tested.  This motivates us to study the LSV  effects on inclusive  production process of heavy flavored hadrons in pair annihilation.  To be specific, we concentrate on the production  of B-mesons (bottom-flavored mesons)  in the process $e^+e^-\to B+jets$ and estimate  the magnitude of symmetry breaking parameters. 	To this aim,   we first calculate the partonic cross section  $e^+e^-\to \gamma/Z\to  q\bar{q}$ to leading order in the perturbative EW theory.  To impose the LSV corrections,  we introduce nonminimal LSV couplings in the EW sector of the SM. Modifications of this sector have been proposed in Refs.~\cite{DeFabritiis:2021rnh,Colladay:2017zen,Colladay:2017qfr,Lehnert:2018lce}, where the modified dispersion relations of the photon and Z-boson were explored. Furthermore, nonminimal LSV couplings were considered in different scenarios within the QED theory, e.g., the scattering processes, the spectrum of hydrogen atom and the electric or magnetic dipole moments \cite{Belich:2006pi,Bolokhov:2006yx,Pospelov:2005pr,deBrito:2016zav}.  Here, we use the LSV prescription from Ref.~\cite{DeFabritiis:2021rnh} where authors have introduced two  (real) Lorentz violation four-vectors, $\rho^\mu$ and $\xi^\mu$, related to the $SU(2)_L$ and $U(1)_Y$ sectors, respectively. This scenario is similar to that analyzed in Refs.~\cite{Mouchrek-Santos:2016upa,Gomes:2019twi,Chen:2016zpw}. In this work,   we single out the case $\rho=\xi$ and study pair annihilation process focusing on the interaction vertices involving gauge bosons and leptons and determine the differential cross section $d\sigma(e^+e^-\to q\bar{q})/d\Omega$ (about the specific choice $\rho=\xi$ we will explain later). In the following, using the fragmentation function of $q/\bar{q}\to B$  we compute the cross section of B-hadron production in pair annihilation. Considering  the experimental data at the collision energies $\sqrt{s}=10.52$~GeV (scale above the $B\bar{B}$ threshold)  and $\sqrt{s} = m_z$ (Z-boson resonance) we are able to place upper bounds (projections) on the LSV parameters, in particular on the time-like component.
	Since, there are few works on scattering processes in the EW sector   including the mentioned nonminimal couplings, then our work could be proposed as a special channel to study the LSV effects.  Our analytical result  for the partonic cross section, which is  presented for the first time,   could be applied to explore  the LSV effects  on the production process of various types of hadrons at the present and future electron-positron colliders such as Circular Electron-Positron Collider (CEPC, in China), Future Circular Collider (FCCee, at CERN), Compact Linear Collider (CLIC, at CERN) and International Linear Collider (ILC, in Japan).

	The remainder of this paper is structured as follows. In Sec.~\ref{sec:three}, the Tree-level cross section for pair annihilation in the presence of vectorial nonminimal couplings is computed. In this section, we analyze the production of hadrons via pair annihilation in the EW theory and estimate the corresponding LSV parameter. Finally, Sec.~\ref{sec:fourr} is devoted to the
	summary and conclusion.

%%%%%%%%%%%%%%%%%%%%%%%%%%%%%%%%%%%%%%%%%%%%%%%%%%%%%%%%%%%%%%
\section{Pair annihilation in the presence of vectorial nonminimal coupling}
\label{sec:three}

At tree level, the differential cross section for the pair annihilation process: $e^-(p_1)+e^+(p_2)\to \gamma/Z/H\to q(p_3)+\bar{q}(p_4)$  is expressed as
\begin{eqnarray}
\frac{d\sigma^{LO}}{d\Omega}=\frac{\overline{|\mathcal{M}|^2}}{64\pi^2 s}\frac{|\vec{p}_3|}{|\vec{p}_1|},
\end{eqnarray}
where,  $s=(p_1+p_2)^2=E_{com}^2$ is the collision energy in the center-of-mass (CM) frame and $|\vec{p}_3|=|\vec{p}_1|\sqrt{1-4m_q^2/s}$.
The pair annihilation process proceeds via the
s-channel exchange of $Z$-boson, photon and Higgs-boson. The corresponding  Feynman diagrams at leading-order (LO) are shown in Fig.~1.  The total Feynman amplitude reads $\mathcal{M}=\mathcal{M}_\gamma+\mathcal{M}_Z+\mathcal{M}_H$. 
It is simple to show that the ratio $\mathcal{M}_H/\mathcal{M}_Z$ is of order $m_em_q(s-m_Z^2)/(m_Z^2(s-m_H^2))$ so that  the contribution $\mathcal{M}_H$ may be neglected unless $s$ is very close to $m_H^2$ (Higgs boson resonance). In this work we neglect the contribution of Higgs boson and compute the mean squared amplitude as $\overline{|\mathcal{M}|^2}=\overline{|\mathcal{M}_\gamma|^2}+\overline{|\mathcal{M}_Z|^2}+2Re(\overline{\mathcal{M}_\gamma\cdot \mathcal{M}_Z^\dagger})$.\\
The  Lagrangian density of the standard Electroweak theory in the unitary gauge is given by
\begin{eqnarray}
\mathcal{L}^{EW}&=&\mathcal L_0+\mathcal L_{I}\nonumber\\
	&&\hspace{-0.5cm}=\bar{\psi}_f(i \displaystyle{\not}\partial -m_f)\psi_f-\frac{1}{4}F_{\mu\nu}F^{\mu\nu} -\frac{1}{4}Z_{\mu\nu}Z^{\mu\nu}\cdots- e\bar{\psi}_f\gamma_\alpha \psi_f A^\alpha-\sqrt{2} \frac{g_W}{\cos\theta_W} \bar{\psi}_f\gamma_{\alpha}(g_V^f-g_A^f \gamma_5)\psi_f Z^\alpha+\cdots,
	\end{eqnarray}
where  $g_V^f=T_3^f-2e_q\sin^2\theta_W$ and $g_A^f=T_3^f$ are the vector coupling and axial-vector coupling (or axial coupling) constants, respectively. Here, $e_q$ stands for the fractional charge of fermions (leptons and quarks) and $\theta_W$ is the weak mixing angle so that one has $sin^2\theta_W=0.227\pm 0.014$. Moreover,  $T_3^f$ is the third component of weak isospin so one has $T_3^f=\pm 1/2$ for  left-handed fermions in $SU(2)_L$ doublets.
%----------------------------------------
\begin{figure}
	\begin{center}
		\includegraphics[width=1\linewidth]{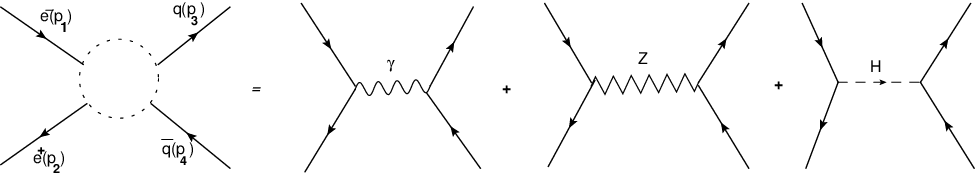}
		\caption{\label{ratio}%
			Feynman diagrams for the process $e^+e^-\to q\bar{q}$ at tree-level.}
	\end{center}
\end{figure}
%-----------------------------------------
Considering the Feynman rules, the amplitude for the photon and $Z$-boson  mediated diagrams are expressed  as
\begin{eqnarray}
	\mathcal{M}_\gamma &=&i\frac{e^2}{s}e_q[\bar{u}_q(p_3)\gamma_{\alpha} v_q(p_4)][\bar{v}_e(p_2)\gamma^\alpha u_e(p_1)],\nonumber\\
		\mathcal{ M}_Z&=&i\frac{2g_W^2}{(s-m_Z^2-im_Z\Gamma_Z)\cos^2\theta_W}[\bar{u}_q(p_3)\gamma_{\alpha}(g_{V}^q-g_{A}^q\gamma_5) v_q(p_4)][\bar{v}_e(p_2)\gamma^\alpha (g_{V}^e-g_{A}^e\gamma_5) u_e(p_1)],
\end{eqnarray}
where, $g_{A}^e=-1/2$ and $g_{V}^e=2\sin^2\theta_W-1/2$ (for $Ze^+e^-$ coupling) and for the $q\bar{q}Z$  coupling $g_{A}^q=T_3^f=\pm 1/2$ (+1/2 for the Up-type quarks and -1/2 for Down-quarks). Also, the mass and decay width of $Z$-boson is set as $m_Z=91.187$~GeV and $\Gamma_Z=2.495$~GeV. 
Before we proceed,  the kinematics of annihilation process are briefly addressed. Considering the kinematics of  process in the CM frame,  the four-vector momenta take the following form
\begin{eqnarray}
	&&p_1=E(1, 0, 0, 1),\quad p_3=(E, p\, sin\theta \cos\phi, p\, sin\theta\sin\phi, p\, \cos\theta)\nonumber\\
	&& p_2=E(1, 0, 0, -1),\quad p_4=(E, -p\, sin\theta \cos\phi, -p\, sin\theta\sin\phi, -p\, \cos\theta) 
\end{eqnarray}
where we assumed the $\hat{Z}$-axis as the collision axis and   $\theta$ is the polar  angle relative to the initial beam direction. Here, we labeled  $|\vec{p}_3|=|\vec{p}_4|=p$ in which $p=\sqrt{s/4-m_q^2}$.
By ignoring the electron mass and taking $g_V^q\equiv g_{qV},\, g_A^q\equiv g_{qA},\, g_V^e\equiv g_{eV},\, g_A^e\equiv g_{eA}$, the LO cross section of $e^+e^-\to q\bar{q}$ in the SM reads
\begin{eqnarray}\label{smm}
	\sigma^{SM}_0(e^+e^-\to q\bar{q})&=&\frac{ N_c}{3\pi s}\sqrt{1-\frac{4m_q^2}{s}}\Bigg\{4\pi^2\alpha^2 e_q^2(1+\frac{2m_q^2}{s})+\frac{1}{(s-m_Z^2)^2+m_Z^2\Gamma_Z^2}\bigg[\frac{s^2g_W^4}{\cos^4\theta_W}(g_{eA}^2+g_{eV}^2)[g_{qA}^2+g_{qV}^2\nonumber\\
	&&-\frac{2m_q^2}{s}(2g_{qA}^2-g_{qV}^2)]+\frac{4\pi\alpha g_W^2e_q}{\cos^2\theta_W} g_{eV}g_{qV}(s-m_Z^2)(s+2m_q^2)\bigg]\Bigg\}.
\end{eqnarray}
where, $(g_W/\cos \theta_w)^2=4\sqrt{2}m_z ^2 G_F$ that $G_F(=1.166\times10^{-5}GeV^{-2})$ is the Fermi constant,  $\alpha$ is the Sommerfeld  constant and $N_c=3$ is the number of quark's colors.  We take $\alpha$ to be the running fine-structure constant so for the scales $\sqrt{s}=10.52$~GeV and $\sqrt{s}=m_Z$ we have $\alpha\approx 1/132$ and $1/128$, respectively \cite{Hagiwara:1994pw,Zhang:2021ypo}. 
If we replace $N_c\to 1$, $e_q\to 1$, $g_{qV} \to g_{eV}$, $g_{qA} \to g_{eA}$ and take $m_q=0$  we achieve to the well-known result for  $\sigma(e^+e^-\to \mu^+\mu^-)$ when $m_\mu=0$ is set \cite{QFT}. In
most references, for simplicity, the massless assumption of fermions is considered. Here, we preserve the quark mass to have more
accurate result. This point is so important when the LSV effects are concerned. To check the b-quark mass dependence of  $\sigma(e^+e^-\to b\bar{b})$  we computed the ratio  $\sigma^{SM}(m_b=0)/\sigma^{SM}(m_b\neq 0)$, considering $m_b=4.9$~GeV, see Fig.~\ref{ratio}. Result shows that the massive assumption  leads to a modification  of  cross section by as much as $1.4\%$  at $\sqrt{s}=m_Z$. This accuracy seems to be important when it is compared with the NLO QCD radiation corrections to the process  $e^+e^-\rightarrow  b\bar b+(g)$ \cite{Chetyrkin:1979bj}, i.e.,
\begin{eqnarray}
	\Big(\frac{\sigma_{NLO}}{\sigma_{LO}}\Big)_{m_b=0}\propto 1+
	\frac{3C_F}{4\pi}\alpha_s(\mu)= 1+0.318 \alpha_s(\mu).
\end{eqnarray}
Taking $\alpha_s^{(n_f=5)}(m_Z)=0.1184$ one has $\sigma_{NLO}/\sigma_{LO}\approx 1.037$ while  $\sigma^{SM}(m_b=0)/\sigma^{SM}(m_b\neq 0)\approx 1.014$  for $\sqrt{s}=m_Z$. The ratio $\sigma^{SM}(m_b=0)/\sigma^{SM}(m_b\neq 0)$ is $1.918$ for $\sqrt{s}=10.52$. Since the data from Belle Collaboration for the inclusive cross section $e^+e^-\to B+X$ has been  reported for $\sqrt{s}> 10.52$~GeV (the scale above the $B\bar{B}$ threshold) \cite{OPAL:1995rqo,Belle:2021lzm} then the b-quark mass effect seems to be important so we preserve it in our calculation.

%----------------------------------------
\begin{figure}
	\begin{center}
		\includegraphics[width=0.49\linewidth]{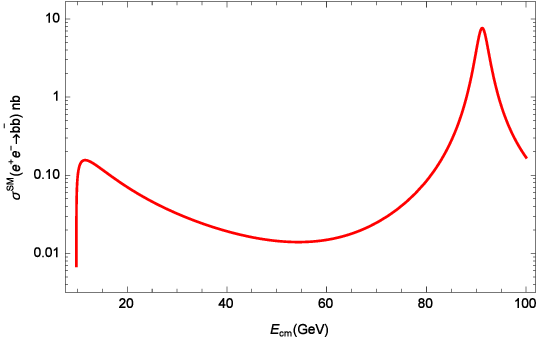}
		\includegraphics[width=0.49\linewidth]{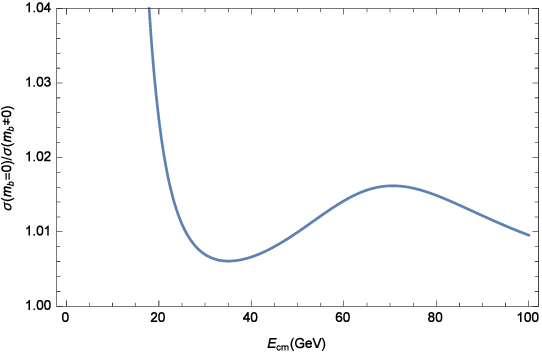}
		\caption{\label{ratio}%
			Left:  $\sigma_0^{SM}(e^+e^-\to b\bar{b})$ for $m_b\neq 0$ on a  logarithmic scale. Right: Cross section for $m_b=0$ normalized to the one for $m_b\ne 0$. }
	\end{center}
\end{figure}
%-----------------------------------------

As was mentioned in the Introduction, an interesting scenario to introduce the Lorentz violation in the EW theory is to modify the gauge boson-fermion vertices directly.   
In this way, for example,  one obtains an extended version of QED characterized by the following  nonminimal covariant derivative \cite{Bakke:2012gt}
\begin{eqnarray}
	D_\mu=\partial_\mu+ieA_\mu+i\mathfrak{g}\eta^\nu F_{\mu\nu}^\star
\end{eqnarray}
where $F_{\mu\nu}^\star$ is the dual electromagnetic field tensor which is related to the usual field-strength tensor as $F_{\mu\nu}^\star=\epsilon_{\mu\nu\alpha\beta}F^{\alpha\beta}/2$ with $\epsilon^{0123}=1$ (four-dimensional  antisymmetric Levi-Civita tensor). Moreover, $e, \mathfrak{g}$ and $\eta^\nu$
are the electron charge, a coupling constant and a constant four-vector, respectively, and the LSV coupling constant has mass dimension $[\mathfrak{g}\eta^\nu]=-1$. In this situation, the additional term sets a nonminimal coupling of the fermion sector to a fixed background $\eta^\nu$, however  it is clearly gauge invariant. The LSV background $\eta^\nu$ is responsible for the breaking of Lorentz symmetry at the particle frame,  as it may select a preferred direction in space-time.   With this modification, the generalized QED Lagrangian (including  Lorentz violating CPT-odd term) in the Feynman gauge is written as
\begin{eqnarray}
	\mathcal{L}_{QED}=\bar{\psi}_f(i\gamma^\mu\partial_\mu-m_f)\psi_f-\frac{1}{4} F^{\mu\nu}F_{\mu\nu}-e\bar{\psi}_f\displaystyle{\not}A \psi_f -(\mathfrak{g}\eta^\nu)\bar{\psi}_f\gamma^\mu \psi_f (\partial^\alpha A^\beta)\epsilon_{\mu\nu\alpha\beta}.
\end{eqnarray}
Through this extended Lagrangian the Feynman rules could be extracted,  as in the standard QED, so that the coupling factors in the photon-electron vertex,  $e^+e^-\gamma(k)$, is modified as: $
	ie\gamma^\mu\to ie\gamma^\mu+\gamma^\mu{\cal O}(\mathfrak{g}\eta\cdot k).$
In  Refs.~\cite{Charneski:2012py,deBrito:2016zav,Ekraminasab:2024vfu}, using this scenario authors have studied the effect of  Lorentz violating nonminimal coupling on the Bhabha and Compton  scatterings as well as pair annihilation in QED. 
Through this work we intend to study  the effect of  LSV nonminimal coupling on the pair annihilation in the Electroweak theory. Lorentz-symmetry violation may be incorporated into
the electroweak sector by considering couplings analogous
to those in the QED sector of the (minimal) SME. 
To achieve, there are various possibilities. One interesting possibility, which was described above and is adopted in our work, is to directly modify the
interaction vertices between mediators and matter fields.
In the SM, the tree-level interactions between leptons $\psi_f$ and gauge bosons  are exclusively derived from the terms $\sim i\bar{\psi}_f \gamma^\mu D_\mu \psi_f$.
As in QED, in the EW theory the leptons are minimally coupled to the
gauge bosons via the following covariant derivative
\begin{eqnarray}\label{zendan}
	D_\mu=\partial_\mu-ig A_\mu^a T^a-ig^\prime Y B_\mu,
\end{eqnarray}
where $Y$ is the weak hypercharge, $T^a$ are the Hermitean generators of SU(2) satisfying $[T^a,T^b]=if^{abc}T^c$, $g^{\prime}$ and $g$ are the respective $U(1)_Y$ and $SU(2)_L$ coupling constants that satisfy the relation $g\sin\theta_W=g^\prime\cos\theta_W=e$ that $e$ is the fundamental electric  charge.  Following Refs.~\cite{Gomes:2019twi,Mouchrek-Santos:2016upa,deBrito:2016zav,Charneski:2012py} we introduce LSV nonminimal couplings via two
real (constant) 4-vectors $\xi^\mu$ and $\rho^\mu$ with canonical dimension of inverse mass. Following Ref.~\cite{DeFabritiis:2021rnh}, we consider the LSV prescription through following derivative 
\begin{eqnarray}\label{bef}
	D_\mu\to \hat{D}_\mu=D_\mu+ig\rho^\nu F_{\mu\nu}^a T^a+ig^\prime\xi^\nu  Y B_{\mu\nu},
\end{eqnarray}
where the field-strength tensors read  $F_{\mu\nu}^a=\partial_\mu A_\nu^a-\partial_\nu A_\mu^a+g f^{abc} A_\mu^b A_\nu^c$ and $B_{\mu\nu}=\partial_\mu B_\nu-\partial_\nu B_\mu$. This prescription introduces nonminimal Lorentz-violating operators in the Higgs sector in a way consistent with the usual gauge transformations, thus preserving the whole gauge structure of the SM. Following Ref.~\cite{DeFabritiis:2021rnh} we consider a more natural scenario where the background LSV vectors couple in an equivalent way with the different sectors of the gauge group, that is, $\rho=\xi$. This choice result in extended versions of the Bianchi identities (in connection with the extended version of the field-strength tensors). 
In our more natural scenario, the background 4-vector $\rho$ simultaneously modifies the weak hypercharge and isospin sectors and induces LSV effects wherever $\hat{D}_\mu$ present.
Note that, the interaction terms from $D_\mu$ (\ref{zendan}) explicitly depend
on the hypercharge and isospin of the leptons but the LSV
terms in Eq.~(\ref{bef}), however, are insensitive to those \cite{Mouchrek-Santos:2018vrn,Belich:2004ng}. Moreover, the mass terms from the Yukawa interactions are not affected since the LSV terms do not influence them
at tree level, meaning that the equations of motion for the
free leptons are unchanged and the propagators—and
associated Feynman rules—will be the same as in the SM.

Considering the extended derivative in Eq.~(\ref{bef}), the interaction Lagrangian containing the LSV background reads:
\begin{eqnarray}\label{mary}
	\mathcal{L}_I^{LSV}=\frac{\sqrt{2}g_W}{\cos\theta_W}\rho^\nu\bar{\psi}_f\gamma^\mu(g_V^f-g_A^f\gamma_5)\psi_f\partial_\nu\hat{Z}_\mu+\cdots,
\end{eqnarray}
where $\hat{Z}^\mu=Z^\mu-\partial^\mu(\rho\cdot Z)$ is a redefinition of the massive Z-field, satisfying the usual subsidiary condition $\partial_\mu \hat{Z}^\mu =0$.  Having the interaction Lagrangian, the Feynman rules governing the lepton-boson interactions can be extracted. \\
Our goal is to calculate the cross section of hadroproduction from pair annihilation which has been experimentally measured and, under the—so far justified—assumption that the SM appropriately describes the central value of the experimental results, use the quoted uncertainties to extract upper limits on the LSV coefficients. To this aim, we first compute the cross section of $e^-(p_1)+e^+(p_2)\to \gamma/Z\to q(p_3)+\bar{q}(p_4)$ at the LSV framework. The coupling factor in the electron-Z vertex ($e^+e^-Z(k)$) may be read directly from Eq.~(\ref{mary}) which already includes the redefined fields. The result is  \cite{DeFabritiis:2021rnh}
\begin{eqnarray}\label{matin}
	i \frac{\sqrt{2}g_W}{\cos{\theta_w}}\gamma^\mu(g_V-g_A \gamma_5)\to i \frac{\sqrt{2}g_W}{\cos{\theta_w}}\gamma^\mu(g_V-g_A \gamma_5)[1+i\rho\cdot  k].
\end{eqnarray}
In a similar way, the modification of electron-photon vertex reads: $
ie\gamma_\mu\to ie\gamma_\mu (1+i\rho^\nu k_\nu)$. Here, $k(=p_1+p_2)$ is the four-momentum carried by the intermediated gauge boson (photon and Z-boson). 
Since, the effects of LSV have not been  observed experimentally, then we may assume that these backgrounds are small relative to currently attainable energy scales, so that we are allowed to keep only terms at leading order in the LSV parameters, and consistently neglect all higher-order terms.  
At last but not least, we would like to mention that the LSV couplings are gauge invariant with a negative canonical dimension ($[\rho^\nu]=-1$), i.e., the
LSV terms are nonrenormalizable. This is a general feature
of such nonminimal couplings and indicates that the
associated Lagrangian is, in fact, only an effective theory.
This will not disturb us here, since we are only dealing with
relatively low-energy processes—exclusively at tree level—
so no divergences are expected.

\begin{figure}
	\begin{center}
		\includegraphics[width=1\linewidth]{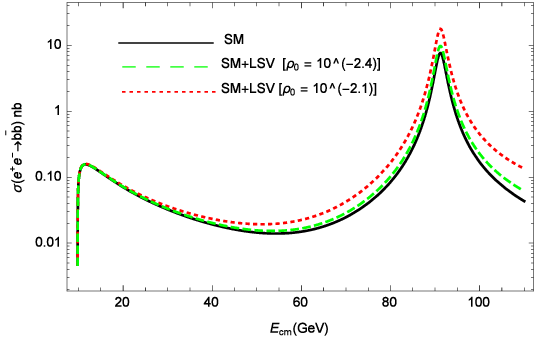}
		\caption{\label{fig4}%
			Total cross section for $e^+e^-\to b\bar{b}$ in the CM framework. Black curve is the SM result and different curves are the combinations of the SM and LSV results for various values of the LSV background, i.e., $\rho_0=10^{-2.4}$ GeV$^{-1}$ for dashed  line and $\rho_0=10^{-2.1}$  GeV$^{-1}$ for dotted. }
	\end{center}
\end{figure}

 For the annihilation process $e^-(p_1)+e^+(p_2)\to \gamma/Z\to q(p_3)+\bar{q}(p_4)$, the corresponding Feynman amplitude at lowest order in the perturbative EW theory is given by
\begin{eqnarray}\label{matin}
	\mathcal{M}=\mathcal{M}_\gamma +\mathcal{M}_Z
	\end{eqnarray}
where:
\begin{eqnarray}
\mathcal{M}_\gamma=\frac{i}{s}\bigg[\bar{u}_q(p_3)\big(iee_q\gamma_\alpha \{1+i\rho^\mu k_\mu\}\big) v_q(p_4)\bigg]\bigg[\bar{v}_e(p_2)\big(ie\gamma^\alpha \{1+i\rho^\nu k_\nu\}\big) u_e(p_1)\bigg],
\end{eqnarray}
and:
\begin{eqnarray}
\mathcal{M}_Z&=&\frac{i}{s-m_Z^2-im_Z\Gamma_Z }\bigg[\bar{u}_q(p_3)\big\{i \frac{\sqrt{2}g_W}{\cos{\theta_w}}\gamma^\alpha(g_{qV}-g_{qA} \gamma_5)[1+i\rho^\mu k_\mu]\big\} v_q(p_4)\bigg]\nonumber\\
&&\hspace{2.5cm}\times \bigg[\bar{v}_e(p_2)\big\{i \frac{\sqrt{2}g_W}{\cos{\theta_w}}\gamma_\alpha(g_{eV}-g_{eA} \gamma_5)[1+i\rho^\nu k_\nu]\big\} u_e(p_1)\bigg].
\end{eqnarray}
 To compute the unpolarized cross section, we need the squared amplitude averaged over the initial spin states and summed over the spin states
\begin{eqnarray}
	\overline{|\mathcal{M}|^2}=\frac{1}{1+2s_e}\frac{1}{1+2s_e}\sum_{spin,color}|\mathcal{M}|^2=\frac{N_c}{4}\sum_{spin}|\mathcal{M}|^2,
\end{eqnarray}
where,  $\overline{|\mathcal{M}|^2}=\overline{|\mathcal{M}_\gamma|^2}+\overline{|\mathcal{M}_Z|^2}+\textbf{Interference term}$.  Using the completeness relations for Dirac spinors, this averaging procedure 
is converted into a trace over Dirac gamma matrix strings. Using the FeynCalc package \cite{Mertig:1990an}, the traces could be computed. As before, we set  $p_1^2 = p_2^2 = m_e^2 \simeq 0$. By integrating over the polar ($0\le \theta\le \pi$) and azimuthal  ($0\le \phi \le 2\pi$) angles the total cross section is obtained as
\begin{eqnarray}\label{bahare}
	\sigma_0^{LSV}(e^+e^-\to q\bar{q})=\sigma_0^{SM}[1+(\rho\cdot k)^2]^2,
\end{eqnarray}
where $\sigma_0^{SM}$ is given in Eq.~(\ref{smm}).
To evaluate the dot-product of gauge bosons four-momentum ($k=p_1+p_2=(2E,\vec{0})$) and the LSV four-vector ($\rho^\mu$), we take the spacelike part of $\rho^\nu$ to point in an arbitrary direction, parametrizing the full four-vector as 
$\rho^\nu =(\rho_0,\vec{\rho})$. 
Then, the scalar product of four-momenta reads: $\rho\cdot k=2E\rho_0=\sqrt{s}\rho_0$ so that one has
\begin{eqnarray}\label{ayda}
	\sigma_0^{LSV}(e^+e^-\to q\bar{q})=\sigma_0^{SM}(1+s\rho_0^2)^2.
\end{eqnarray}
As is seen, no first-order LSV contribution appears. It is also
worthwhile noticing that the LSV piece depends only on
the isotropic time component $\rho_0$ (a purely time-like background) and the LSV-free part reproduces the SM result, as it should, see Refs.~\cite{DeFabritiis:2021rnh,Gomes:2019twi,Chen:2016zpw} where the effect of LSV is tested via electroweak decays. In Fig.~\ref{fig4}, taking $m_b=4.9$~GeV the LSV cross section $\sigma_0^{LV}(e^+e^-\to b\bar{b})$ is shown for different values of the LSV background and for the SM.
%----------------------------------------
%-----------------------------------------

Having the LSV-modified cross section at the parton level, we may discuss possible limits on our time-like background using the experimental data for the process $e^+e^-\to  Hadron+Jets$.
To be specific, we  consider the inclusive pair annihilation process to produce B-hadron; $e^{+}e^{-}\to q\bar{q}\to  B+X$
where, $X$ shows unobserved jets. For studying the  hadron production process in the perturbative Electroweak theory, the factorization theorem of improved parton model plays the key role \cite{Collins:1998rz}. 
Based on this theorem the  cross section of hadron production can be expressed as a sum of convolutions of hard scattering result  $d\sigma_q(y, \mu_R, \mu_F)/dy$    with the
nonperturbative fragmentation functions (FFs) of B-meson from the initial parton $q(=u,\bar u, \cdots, b,\bar b)$, i.e., $D^B_q (x, \mu_F)$. It reads
\begin{eqnarray}
	\label{sigx}
	\frac{d\sigma_B}{dz}=
	\sum_{q}\int\limits_z^1\frac{dy}{y}
	\frac{d\sigma_q}{dy}\left(y, \mu_R,  \mu_F\right)D_q^B(\frac{z}{y}, \mu_F),\qquad q=g,\,u/\bar{u},\,\cdots,b/\bar{b}.
\end{eqnarray}
Here,  the scales $\mu_R$ and $\mu_F$ show the renormalization  and factorization  scales, respectively, so we conventionally adopt $\mu_R^2=\mu_F^2=s$. If we label the four-momenta of B-meson and the intermediate 
gauge photons ($\gamma/Z$) by $p_B$ and $k$, respectively, so that $s=k^2$, then  the scaling variable $z$
is defined as $z=2(p_B.k)/k^2$ so that $0\le z\le 1$. In the CM frame, it is simplified to  $z=2E_B/\sqrt{s}$ which, in fact, refers
to the B-meson energy  scaled to the beam energy. In the CM  frame, the variable $y$ is also defined as $y=2E_q/\sqrt{s}$.
The fragmentation function $D_q^B(x=z/y,\mu_F)$ indicates the probability density of B-meson production from the initial parton $q$ which carries the energy fraction $x(=E_B/E_q)$ of the parent quark.  
 \\
Having the LSV cross section (\ref{ayda}), the LO differential cross section at the parton level  is given by: $d\sigma_0^{LV}(e^+e^-\to q\bar{q})/dy=\sigma^{LV}_0\delta(1-y)$.
By substituting  this result into Eq.~(\ref{sigx}) one gets
\begin{eqnarray}\label{sigxx}
	\frac{d\sigma_B}{dz}(e^+e^-\to  B+jets)=2 \sigma^{LV}_0 \sum_q D_q^B(z, \mu_F),
\end{eqnarray}
where  we assumed $D_q^B=D_{\bar q}^B$. The primary products of $b/\bar{b}$-hadronization are the bottom-flavored mesons $B^+, B^0
, B^0_s$  and their charge conjugates.
Since, the most probability density for production of  a specific meson  comes from the fragmentation of a quark which is itself a  constituent quark of desired meson then the splitting contribution of $b\to B$  is governed, specifically at Tree level. In higher orders the contribution of other partons, including gluon, appears. Then, the relation (\ref{sigxx}) for the inclusive process $e^+e^-\to BX$ is estimated as
\begin{eqnarray}
	\frac{d\sigma_B}{dz}(e^+e^-\to B+X)\approx 2 \sigma^{LV}_0  D_b^B(z, \mu_F).
\end{eqnarray}
Then, the total cross section reads
\begin{eqnarray}\label{fin}
	\sigma_B(e^+e^-\to BX)\approx 2 \sigma^{LV}_0 Br(b\to B),
\end{eqnarray}
where $Br(b\to B)=\int_0^1 dz D_b^B(z, \mu_F)$ represents the branching fraction for the transition $b\to  B$. \\
In Ref.~\cite{Salajegheh:2019ach}, the nonperturbative FF of $b\to B$ at next-to-next-to-leading order (NNLO)  is obtained through a global fit to $e^+e^-\to B+Jets$ annihilation  data from the ALEPH, DELPHI, and OPAL collaborations at CERN LEP1 and the SLD collaboration at SLAC SLC. 
Specifically, the power ansatz $D_b^B(z,\mu_0)=Nz^\alpha(1-z)^\beta$ was used as the initial condition for the $b\to B$ FF at $\mu_0=4.5$~GeV, while the gluon and light-quark FFs were generated via the DGLAP evolution equations \cite{dglap}. The fit yielded $N =1805.896, \alpha=14.168$, and $\beta=2.341$ with $\chi^2/(d.o.f)=1.104$. 
Having the partonic cross section (Eq.~(\ref{ayda})) and $D_b^B$-FF, the theoretical cross section of B-hadron production could be  computed. 
Considering the experimental data for the inclusive production cross section of $B$-mesons  through pair annihilation, reported by the Belle and OPAL Collaborations \cite{OPAL:1995rqo,Belle:2021lzm}, one can approximate  the  time-like component of background  4-vector $\rho$ which is a free parameter in Eq.~(\ref{ayda}).
To determine the value of  $\rho_0$  we consider the size of cross section at the energy scale  $E_{cm}=10.52$~GeV which  is just above threshold for B-meson production and $\sqrt{s}=m_Z$ (Z-boson resonance).   For the CM energy $E_{cm}=10.52$~GeV, the cross section has been reported as $\sigma=3.27\pm 0.004\pm 0.032$~nb (the first error is statistical
and the second includes systematic uncertainties).   Considering the theoretical cross section (\ref{fin}) and the unit conversion $1$~GeV$^{-2}=0.389379\times 10^6$~nb we obtain $\rho_0\approx 35.32\times 10^{-3}$~GeV$^{-1}$. For the CM energy $E_{cm}=m_Z$, the cross section has been reported as $\sigma=41.7\pm 0.043$~nb so one obtains $\rho_0\approx 8.28\times 10^{-3}$~GeV$^{-1}$. These achievements  are consistent  with the value $\le 10^{-12}$~eV$^{-1}$ reported in Ref.~\cite{Charneski:2012py} and  $\le 10^{-3}$~GeV$^{-1}$  in Ref.~\cite{deBrito:2016zav} where  the effect of  Lorentz violating nonminimal coupling on the Bhabha  and   the Compton  scatterings have been studied in the QED theory. We also checked our result with the Data Tables for Lorentz and CPT Violation  \cite{Kostelecky:2008ts} and found good agreement with the one  presented in Table D26-part 3 (Nonminimal  couplings, $d \ge 5$), i.e., $\rho_0 \equiv\xi<10^{-3}$~GeV$^{-1}$ (the LV background $\xi$ is a nonminimal coupling with
canonical dimensions of inverse mass, as in our work.).

In above discussion, we supposed a way to obtain the value of LSV parameter from the analyses of B-meson production  in the EW theory. In the SM, the total partonic  cross section in higher-order approximations is expressed as \cite{Chetyrkin:1979bj}:
\begin{eqnarray}
	\frac{\sigma_{tot}}{\sigma_0}(e^+e^-\to q\bar{q})\approx 1+\frac{\alpha_s}{\pi}+1.411 \alpha_s^2+\cdots ,
\end{eqnarray} 
where, $\alpha_s$ is the QCD coupling constant and $\sigma_0=4\pi\alpha^2/(3s)$ is the cross section of $e^+e^-\to \mu^-\mu^+$ at LO when one sets $m_e=m_\mu=0$. The inclusion of QCD effects would improve the value of  $\rho_0$  and should allow a better comparison with the results encountered for atomic clocks or torsion balances.

As a final and necessary remark, we would like to emphasis a point regarding the components of the LSV four-vector as measured at an Earth-based laboratory. By definition, the background 4-vector $\rho$ is non-dynamical, being fixed in space-time. However, this is manifest only in a truly inertial reference frame. A common and convenient choice is the Sun-centered frame (SCF) \cite{ref11}, which may be considered approximately inertial for our purposes. 
According to Refs.~\cite{Kostelecky:2008ts,Bluhm:2003un}, the axes in the SCF are defined such that the $\hat{Z}$-axis is directed parallel to Earth's
rotational axis, $\hat{X}$ points from the Sun to the vernal equinox,
while $\hat{Y}$ completes a right-handed system; the origin of time
$T$ is at the 2000 vernal equinox. 
Consequently, in Earth-bound experiments, the LSV-modified observables, such as cross sections, should display sidereal time modulations that could be picked up by carefully time-stamping the data. It is thus necessary to express the laboratory-frame components, $\rho_{lab}^\mu$, in terms of the components $\rho_{SCF}^\mu$ which are static in the SCF.
Following Ref.~\cite{deBrito:2016zav}, we find that, up to first order in Earth's orbital velocity ($\sim 10^{-4}$ in units of $c$), $\rho_{lab}^0\approx \rho_{SCF}^T$  while  the spatial components are different in both frames, being related by a time-dependent rotation matrix $R^{iJ}(\chi,T)$ via $\rho_{lab}^i\approx R^{iJ}(\chi,T)\rho_{SCF}^J$. The rotation matrix  is  \cite{Gomes:2019twi}
\[
\begin{bmatrix}
	c_\chi c_{\bigoplus}\, & c_\chi s_{\bigoplus}\, & -s_\chi \\
	-s_\chi\, & c_\chi\, &  0 \\
	s_\chi c_{\bigoplus}\, &  s_\chi s_{\bigoplus}\, & c_\chi 
\end{bmatrix}
\]
where $c_\chi=\cos\chi, s_\chi=\sin\chi, s_{\bigoplus}=\sin(\omega_{\bigoplus}T_{\bigoplus})$ and $c_{\bigoplus}=\cos(\omega_{\bigoplus}T_{\bigoplus})$. Here, $\chi$ is the experiment's co-latitude,  $\omega_{\bigoplus}(=2\pi/day\approx 7\times 10^{-5} s^{-1})$ is the Earth's rotation angular velocity (the sidereal frequency) and the local time $T_{\bigoplus}$ is defined to be the time measured in the SCF from one of
the moments when the axis $\hat{y}$ (in the standard
Earth-bound frame) lies along the axis $\hat{Y}$ in the SCF.
\\
 In the context of the approximations employed, the time component is easily factored out from the squared amplitudes. Our results showed  that the LSV background appears purely time-like at the annihilation process, so this simplified the discussion leading to experimental bounds. Note that, as was mentioned, the time component is approximately identical in the SCF and in the standard Lab frame. Given that the latter and the CM frame are related by a time-independent (spatial) rotation, the time component is the same in all three frames, i.e., the Lab-,  the SCF- and the CM-frames.

\section{Summary and Conclusion}
\label{sec:fourr}

Collision experiments in high energy current and future colliders  provide a suitable environment where Lorentz symmetry breaking can be tested.  In the present work we, for the first time, studied  the implications of this symmetry breaking on B-meson production through pair annihilation in the framework of Electroweak theory. We studied a modification to the Glashow-Salam-Weinberg electroweak model through nonminimal couplings
in the non-Abelian and Abelian sectors of the lepton-bosons
interaction. These couplings introduce LSV via two 
real 4-vectors that give rise to a preferred
orientation in space-time. In our more natural scenario, the background 4-vector $\rho$ simultaneously modifies the weak hypercharge and isospin sectors and induces LSV effects wherever the covariant derivative $D_\mu$ present. We first  computed the $d\sigma(e^+e^-\to q\bar{q})/d\Omega$ in the LSV framework at Tree-level and noted that the leading-order LSV contribution is of ${\cal O}(\rho^2)$, a fact also encountered in other works \cite{Charneski:2012py,deBrito:2016zav,Mouchrek-Santos:2016upa,Chen:2016zpw,Gomes:2019twi}. From Eq.~(\ref{ayda}) we obtained sensitivity for the case of a purely time-like LSV background.
In the following, taking  the nonperturbative $D_b^B(z, \mu_F)$-FF from phenomenological works we computed the LO cross section of $\sigma(e^+e^-\to B+jets)$ and compared   with the experimental data taken from the OPAL and Belle Collaborations. From this comparison, we obtained the  magnitude of  corresponding LSV coefficient ($\rho_0\le 10^{-3}$)  which is in good consistency with other analysis resulted from different processes. 
Since,  there is no mathematical formalism or physical principle  which can constrain the magnitude order of the  LSV coefficients, thus we hope that our results may be useful as a guide in the investigation of  Lorentz violation phenomena in high energy scattering processes, especially in next-generation $e^+e^-$ colliders reaching center-of-mass energies at the TeV scale.
These colliders will probe the SM with unprecedented precision and provide an outstanding opportunity to search for possible deviations. 
It should be noted that,  in the present work we applied the fragmentation function of $D_b^B(z, \mu)$ to investigate the B-meson production in pair annihilation but it would also be possible to study the production processes of other hadrons considering their corresponding FFs. For example, in Refs.~\cite{Salajegheh:2019srg,Delpasand:2020vlb,Delpasand:2019xpk,MohammadMoosaviNejad:2018pyp,Delpasand:2020mqv}, the FFs of $D$- and $\eta$-mesons as well as  $\Lambda_c$- and $\Omega_c$-baryons have been calculated up to next-to-next-to-leading order (NNLO) in the theory of perturbative QCD. These processes have enough potential to be tested in the present and  future $e^+e^-$ colliders  such as the international linear collider (ILC)  and  circular electron-positron collider (CEPC). The CEPC is initially planned to have a collision energy of $500$~GeV  with the possibility for a later upgrade to $1$~TeV. 
Up to now, most studies on the production of doubly heavy meson $B_c$  have concentrated on the  hadronic production due to the numerous production rate at high energy colliders such as CERN LHC or Tevatron \cite{Berezhnoy:2019yei,Chen:2018obq,Chang:2005wd}.   The  CEPC machine can be a potentially good platform for investigating the $B_c$ meson. In comparison with high energy hadron colliders there will be less backgrounds at the $e^+e^-$  collider. Thus, it would be suitable for studying this doubly heavy meson, precisely. Furthermore,  the production mechanism of  $B_c$  at the $e^+e^-$ collider is much simpler than the one at hadron colliders. Therefore, our analytical results  as well as our proposed channel could be applicable in the present and future generation of high energy $e^+e^-$ colliders.

$\textbf {Declarations}$\\
$\textbf { Funding}$ There is no funding for this manuscript.\\
$\textbf {Availability of data and material}$ Used data could be found in the references introduced.\\
$\textbf {Competing interests}$ The authors declare that they have no conflict of interest.\\
$\textbf { Authors Contributions}$ V.~Ekraminasab: Conceptualization; Formal analysis; Methodology; Writing-Original Draft. S.~M.~Moosavi Nejad: Investigation; Validation; and Writing-Review.

\end{document}